# Self-oscillation conditions of a resonant-nano-electromechanical mass sensor

Eric Colinet[+], Laurent Duraffourg[+], Sébastien Labarthe[+,*], Philippe Andreucci[+], Sébastien Hentz[+], and Philippe Robert[+]

[+]CEA-LETI MINATEC, 17 rue des Martyrs 38054 GRENOBLE Cedex 09 – France

* Institut Néel, CNRS and Université Joseph Fourier, BP 166, F-38042 Grenoble Cedex 09, France

This article presents a comprehensive study and design methodology of co-integrated oscillators for nano mass sensing application based on resonant Nano-Electro-Mechanical-System (NEMS). In particular, it reports the capacitive with the piezoresistive transduction schemes in terms of the overall sensor performance. The developed model is clearly in accordance with the general experimental observations obtained for NEMS-based mass detection. The piezoresistive devices are much sensitive (up to 10 zg/√Hz) than capacitive ones (close to 100 zg/√Hz) since they can work at higher frequency. Moreover, the high doped silicon piezoresistive gauge, which is of a great interest for very large scale integration displays similar theoretical resolution than the metallic gauge already used experimentally.

## Introduction

In the last several years, nano-electromechanical systems (NEMS) resonators have shown their strong potential as highly resolved nano-sensor for gas sensing and ultra low mass detection applications [1], [2], [3]. In these resonant applications, the adsorbed mass on top of the NEMS induces a shift of its resonant frequency that is continuously measured. To follow this frequency variation, the NEMS is embedded in a closed loop that can be either a phase-locked loop (PLL) or a self-oscillating loop [1]. In such architectures, the mechanical amplitude of the vibrating structure is usually kept constant to ensure the stability of the system. Our work is focused on the self-excited scheme because it is a compact co-integrated mass sensor system. Self-oscillating state is established if the electronic gain supplied by the loop is large enough to compensate the losses of the measurement chain. The electronics must also compensate the phase shift induced by the NEMS.

The mass-resolution of such a device is limited by different noise processes such as thermomechanical, Nyquist–Johnson, or adsorption–desorption noises. It is also limited by the electronics sensitivity. Other noise sources as thermal fluctuations and defect motion-induced noise can also be relevant when the targeted ultimate mass is close to some hundred Daltons.

The stability of the self-excited architecture must then be carefully studied to know the expected performance of the resonant mass sensor. Basically, the needed electronic gain to reach the stable state of the device depends strongly on the transduction technique. In this article, a theoretical study of a NEMS embedded in a self-excited loop with capacitive [2] [4], or piezoresistive (PZR) [5] detection principles is presented. Field effect transduction [6] that may also be used can be treated in the same manner as the capacitive detection.

The same NEMS geometry is used to estimate the performance of the two detection principles. There are many factors that determine the mass sensitivity for a given detection. To make a comprehensive comparison, the NEMS is interfaced at its optimal operating point, e.g. maximum bias voltage taking into account to the physical limitations (pull-in for the capacitive approach, self-heating of the gauges for the piezoresistive way).

The NEMS is based on a vibrating nano cantilever moving in the wafer plane [7]. The cantilever is actuated with lateral electrodes. To perform the PZR detection, gauges are placed on each sides of the cantilever close to the anchor. The gauge are then strained or extended when the beam vibrates. For the capacitive detection is achieved with a second electrode.

The article is organized as follows: The first part the oscillator general topology is presented. The practical implementation of the mechanical structure is then detailed. This structure can be used in capacitive detection or in piezoresistive detection as well. The second part deals with the so-called “capacitive oscillator” where the detection and the actuation of the resonant NEMS are purely electrostatic. In the third part, the “piezoresistive oscillator” based on electrostatic actuation and piezoresistive detection is presented. Finally, ultimate performance (mass resolution) of each oscillator is evaluated.

## General topology of an oscillator

Whatever the transduction principle, the oscillator can be described by the general topology shown in Fig 1. The mechanical resonator displacement is converted into an electrical signal by an impedance variation *ΔZ* through either capacitive or PZR detection. This signal is amplified and phase shifted to fullfil the Barkhausen condition [8] and then feedbacked to the NEMS through the electrostatic actuation.

### Oscillation conditions

To realize a NEMS-based oscillator, it is critical to adjust the open-loop gain and phase changes to satisfy the Barkhausen criterion on resonant $\Omega_1$,

$$H_{OL}(j\Omega_1) = \frac{Vout}{Vin}(j\Omega_1) = H(j\Omega_1)G(j\Omega_1) = 1 \Leftrightarrow \begin{cases} \left\| H_{OL}(j\Omega_1) \right\| = 1 \\ \arg\left(H_{OL}(j\Omega_1)\right) = 0 \end{cases} \quad (1)$$

where $H_{OL}$ is the overall transfer function in open loop configuration. $H(j\omega) = J\alpha(j\omega)\kappa$ is the transfer function of the NEMS. $J$, $\alpha(j\omega)$ and $\kappa$ are the detection gain, the NEMS dynamic and the actuation gain respectively. $G(j\omega)$ is the electronic transfer function to be adjusted.

Whatever the mechanical resonator considered (beam, nanowire, disk…), its dynamical displacement can be reduced to an equivalent mass/spring/damper scheme as follows [9],

$$\alpha(j\omega_0) = \frac{1}{-M_{eff}\omega^2 + K_{eff} + j\omega B_{eff}} = \frac{1/M_{eff}}{-\omega^2 + \Omega_1^2 + j\omega\Omega_1/Q} \quad (2)$$

where $M_{eff}$, $K_{eff}$ and $B_{eff}$ are the effective mass, the effective stiffness and the effective damping coefficient respectively. The values of $M_{eff}$, $K_{eff}$, which depend on both NEMS geometrical parameters and anchorage conditions, define the eigen free frequency $\Omega_1 = \sqrt{K_{eff} / M_{eff}}$ . $B_{eff}$ is related to the quality factor $Q=M_{eff}\Omega_1/B_{eff}$. For NEMS Q is typically close to 1000 in vacuum and 100 in air.

Using (2), (1) is reduced the following expression,

$$\begin{cases} \left\| H_{OL}(j\Omega_1) \right\| = 1 \\ \arg\left(H_{OL}(j\Omega_1)\right) = 0 \end{cases} \Leftrightarrow \begin{cases} \left\| G(j\Omega_1) \right\| = \dfrac{M_{eff}\Omega_1^2}{\kappa JQ} \\ \arg\left(G(j\Omega_1)\right) = -90° \end{cases} \tag{3}$$

where the product $\kappa J$ is considered positive.

**Phase noise**

The oscillator stability is characterized by its frequency noise $\sigma_\omega$ i.e. the frequency fluctuation around its eigen frequency $\Omega_1$. $\sigma_\omega$ is related to the frequency noise spectral density $S_\omega$ via the Parseval-Plancherel theorem,

$$\sigma_\omega^2 = S_\omega(\omega = \Omega_1)\Delta\omega \tag{4}$$

where $\Delta\omega$ is the measurement bandwidth. The maximal bandwidth is set by the NEMS bandwidth $\frac{\Omega_1}{Q}$. To increase the performance, the signal must be integrated over several periods to have for instance 1 Hz-bandwidth resulting in a relative low response time.

From Robins's expression, the frequency noise spectral density is function of the open loop noise spectral density $S_{openloop}^{y_n}(\omega)$ [10], [11],

$$S_\omega(\omega) \approx \frac{\Omega_1^2 S_{openloop}^{y_n}(\omega)}{4Q^2 y^2} \tag{5}$$

$S^{y_n}_{openloop}(\omega)$ is the sum of all noise sources (white thermomechanical noise, Johnson noise, electronic noise) referred to the resonator mechanical displacement. *y* is the RMS value of the beam deflection, which can be controlled through electrical non linearity (e.g. saturation of the electronic gain or by controlling the loop gain).

Considering that the noises are not correlated the overall noise density of the open loop is the sum all noise densities,

$$S^{y_n}_{openloop}(\omega = \Omega_1) = S^{y_n}_{Thermomechanical}(\Omega_1) + S^{y_n}_{Johnson}(\Omega_1) + S^{y_n}_{Electronics}(\Omega_1) \quad (6)$$

The mathematical expressions of each noise density have to be defined according to the transduction principle and will be described further in the next part. The equation (5) shows that the overall noise density is minimized for large displacement *y*. In other words, the dissipated power in the vibrating beam must be larger as possible to limit the frequency noise.

The mass resolution (minimum mass detectable) $\sigma_M$ can be deduced from the phase noise,

$$\sigma_M = 2M_{eff}\frac{\sigma_\omega}{\Omega_1} \quad (7)$$

The effective mass must be low and the resonant frequency must be as high as possible. These characteristics are fulfilled by NEMS structures.

## Practical implementation of the NEMS oscillator

### Architecture

The considered structure is a cross beam as shown in Fig. 2. Table I gives the typical feature sizes of the device. This beam used as a lever arm is actuated by the electrostatic force. The detection of its motion is performed either by two perpendicular thin piezoresistive gauges or by the capacitive variation between the beam and a second electrode. This architecture is versatile and can be used in both configurations. The geometric parameters and the material properties remain the same in both configurations. According to equation (3) $\kappa$ and $J$ have to be defined for each detection case. The electronics transfer function will be designed according to $\kappa$ and $J$. The electronics is based on the same Pierce scheme as shown in Fig.3.

**Transduction modeling**

Second Newton equation is applied in the Galilean referential frame of the substrate to the cross-beam. The cross-beam is assumed to follow Euler-Bernoulli equation [12]. The equation is reduced to its normalized lumped expansion on the first mode through the Galerkin method [13]. Here, we only consider the first eigen modal vector $\chi_1(x)$, which observes the boundary conditions $\chi_1'(-l_1) = \chi_1(-l_1) = 0$ and $\chi'''_1(l - l_1) = \chi_1''(l - l_1) = 0$. $\chi_1(x)$ must also follow the mechanical continuity at the gauge position $\chi_1'(0^-) = \chi_1'(0^+)$ and $\chi_1''(0^-) = \chi_1''(0^+)$. The equivalent electromechanical response of the cross-beam in the "spring/mass/damper" model is expressed as,

$$\alpha(\omega) = \frac{1}{M_{eff}\,(\Omega_1^2 - \omega^2 - j\,\frac{\omega\Omega_1}{Q})} \tag{8}$$

where $M_{eff} = \frac{a}{\chi_1(l - l_1)} \frac{\rho Sl}{\eta_1}$ is the effective mass of the resonant beam for the first eigen mode $\Omega_1 = \sqrt{\frac{EI}{\rho S}} k_1^2$. $Q = \frac{\rho S \Omega_1}{b}$ is the quality factor of the structure ($b$ is the damping coefficient). $\eta_1 = \frac{1}{l^2}\int_{l-l_1-a}^{l-l_1} \chi_1(x)dx$ is a constant depending on the modal base and on the force repartition along the beam. *S*, $\rho$, *E* and *I* are the section of the cross-beam, the density of the silicon (2330 kg/m$^3$), the Young modulus (169 GPa) and the quadratic moment respectively. $\chi_1(l-l_1)$ is the value of the modal function at the beam end. $k_1$ is the wave vector that depends on the distance $l_1$ between the anchor of the lever beam and the gauge. This value has been set to enhance the stress inside the gauge sensors due to a mechanical lever effect and was fixed at $l_1=0.15l$. In that case $k_1=2.12/l$.

The actuation is made through an electrostatic force, which is uniform along the electrode. Within the limit of small deflection compared to the gap *g* and for AC-voltages $V_{ac}$ smaller than the DC-voltage $V_{dc}$, the force per unit length at $\omega=\Omega_1$ is,

$$f(t) \approx \frac{\varepsilon_0 e a V_{dc} V_{ac}}{g^2}\cos(\Omega_1 t) \qquad (9)$$

where $\varepsilon_0$ is the vacuum permittivity (8.85x10$^{-12}$ F.m$^{-1}$).

From (9), the actuation gain can be written as follows,

$$J = \frac{\varepsilon_0 e a V_{dc}}{g^2} \qquad (10)$$

Criterion (3) can now be studied for each detection principle.

## Capacitive oscillator

### Detection gain

The displacement $y(\omega)=\alpha(\omega)F(\omega)$ implies a capacitance variation between the lever arm and the electrode. $|y(\omega)|$ is the displacement of the cross-beam end in the Fourier space. The detection gain corresponding to the conversion of the displacement *y* to the capacitance variation $\delta C$ is then calculated as,

$$\kappa = \frac{\delta C(\omega)}{y(\omega)} = \frac{\varepsilon_0 el^2 \eta_1}{g^2 \chi (l - l_1)} \tag{11}$$

**Circuit modeling**

The figure 4 shows the equivalent small signal circuit of the ensemble variable capacitance and its electronics (see Fig 3.b). The electronics is reduced to an equivalent MOSFET transistor. The input capacitance is the gate/source capacitance in parallel with a certain coupling capacitance. The feedback capacitance is the gate/drain capacitance in parallel with the capacitance due to the actuation capacitance. δC (ΔZ in Fig.1) is the capacitance variation around its nominal value $C_0 = \varepsilon_0 el / g$ $(el >> g^2)$. The fringe effect can be included using the results presented in Ref [14] and will increase the absolute value of some units. The electronic gain using Millman's theorem are defined at nodes 1 and 2,

$$V_1 = \frac{V_{dc}\, j\delta C\omega + V_2\, jC_{fb}\omega}{j\omega(C_{fb} + C_{in} + C_0)} \tag{12}$$

$$V_2 = \frac{(jC_{fb}\omega - g_m)V_1}{j\omega(C_{fb} + C_L) + g_{ds} + 1/R_L} \tag{13}$$

At this stage some assumptions on the relative values of $R_L$, $C_L$, $C_{fb}$, and $C_{in}$ have to be made. In particular, the input current must flow through the capacitance $C_{in}$ and the output current must flow through the capacitance $C_L$,

$$\Omega_1 >> \Omega_L = \frac{1}{C_L (R_L / 1 + R_L g_{ds})} \text{ and } g_m C_{fb} << C_L (C_{fb} + C_{in} + C_0)\Omega_1 \tag{14}$$

where $\Omega_L$ is the cut-off frequency of the electronics.

From (13) and (14), one can deduce,

$$V_{out} \equiv V_2 \approx \frac{-g_m V_1}{j\omega C_L} \tag{15}$$

Using (12) and (15), $G(\omega)$ is,

$$G(\omega) = \frac{V_2(\omega)}{\delta C(\omega)} = \frac{-g_m V_{dc}}{jC_L(C_{fb} + C_{in} + C_0)\omega} \tag{16}$$

The minimal transconductance $g_m$ is defined from the oscillation conditions (3), and the expression (16),

$$g_m \approx \frac{C_L(C_{fb} + C_{in} + C_0)M_{eff}\Omega_1^3}{V_{dc}J\kappa Q} = \frac{M_{eff}C_L\Omega_1^3 g^4 (C_{fb} + C_{in} + C_0)\chi(l - l_1)}{QV_{dc}^2 \varepsilon_0^2 e^2 l^3 \eta_1} \tag{17}$$

The conditions (14) can be simplified using (17) and expressing the conductance $g_{ds}$ as a linear function of $g_m$. Considering a MOSFET in saturation regime it is known that,

$$g_{ds} \approx \frac{|V_G - V_{th}| g_m}{2V_E L} = \beta g_m \tag{18}$$

$V_G$, $V_{th}$ $V_E$ and $L$ are the gate voltage, the threshold voltage, the Early voltage, and the channel length of the equivalent MOSFET respectively. For the 0.35 µm-CMOS technology, their typical values are 3.3V, 0.485V, 21V/µm and 5 µm respectively.

Finally, the conditions (14) are,

$$\frac{1}{R_L C_L \Omega_1} + \frac{\beta\Omega_1^2 M_{eff}(C_{fb} + C_{in} + C_0)}{V_{dc}J\kappa Q} << 1, \forall\Omega_1 \tag{19}$$

and $$\frac{C_{fb}}{\frac{V_{read}J\kappa Q}{M_{eff}\Omega_1^2}} << 1 \tag{20}$$

Notice that the parasitic capacitances included in $C_{in}$ have to be minimized to optimize detection gain. The feedback capacitance is also a strong source of perturbation. $C_{fb}$ reduces the frequency range of the operation.

### Overall noise

To evaluate the mass resolution, the noise sources must be considered (see equation (5)). Here, the white thermomechanical noise and the electronics noise are estimated in a very straightforward manner.

Hence in ($m^2$/Hz)

$$S^{y_n}_{Thermomechanical}(\omega) = \frac{4k_B TQ}{M_{eff}\Omega_1^3} \qquad (21)$$

From the equations (12) and (14), the noise density of the electronic (in $m^2$/Hz) is as follows,

$$S^{y_n}_{Electronics}(\omega) = \frac{1}{V_{dc}^2 |\kappa|^2}\left((C_{fb} + C_{in} + C_0)^2 + \frac{C_{fb}^2 g_m^2}{\Omega_1^2 C_L^2}\right)\left(\frac{8k_B T}{3g_m}\right) \qquad (22)$$

$\left(\frac{8k_B T}{3g_m}\right)$ is the voltage noise density at the input of the electronics.

Regardless of the NEMS geometry, the critical parameters to reach self-oscillations (16) and optimal mass resolution (5), (22) are the DC-voltage, the gap, the capacitances and the deflection amplitude. Discussion on the system performance is done in the last part of the article.

## Piezoresistive oscillator

### Detection gain

The well known piezoresitive equation for the axial stress $\sigma_L$ is [15],

$$\frac{dR}{R} = \frac{d\delta}{\delta} + \varepsilon_L(1+2\nu) \qquad (23)$$

$$\frac{d\delta}{\delta} = \Pi_L \sigma_L = \Pi_L E \varepsilon_L \tag{24}$$

$\Pi_L$ is the longitudinal piezoresistive coefficient along the gauge. Substituting (24) into (23), we get the usual expression,

$$\frac{dR}{R} = (1 + 2\nu)\varepsilon_L + \Pi_L E \varepsilon_L = \gamma_L \varepsilon_L \tag{25}$$

$\gamma_L$ is the gauge factor.

The first part of (25) corresponds to the gauge elongation. The second one takes into account the intrinsic piezoresistive coefficient that is predominant for semiconductor gauge. Its large value permits to attain strong sensitivity. The drawback is its large resistance due to the small section of the gauge that is more or less doped (over a range of $10^{15}$ $cm^{-3}$ to $10^{19}$ $cm^{-3}$). This transduction might not be suitable for very tiny mass detection because of the strong Johnson noise floor. A metallic layers can also be used as the gauge layers (for instance coated thin layer on the silicon beam). In this case the elongation effect is only used. Its advantage is clearly the low resistance with a very low Johnson noise [16]. However, the gauge factor is only a function of the Poisson ratio and its value is reduced to some units.

According to equation (24), the relative resistance variation is proportional to the axial stress $\sigma_L$ exerted by the resonant lever beam. It is given by the difference of shear forces $V^- = V(x = 0^-)$ and $V^+ = V(x = 0^+)$. After few mathematical developments, $\sigma_L$ can be expressed as,

$$\sigma_L = \frac{V^- - V^+}{e w_j} = \frac{\vartheta E I k_1^3}{s} y \tag{26}$$

$s$ is the section of the gauges. $\vartheta$ is a constant that depends on the boundary and normalization conditions.

The displacement $y$ induces the axial stress and implies a relative resistance variation leading to the PZR detection gain,

$$\kappa = \frac{\delta R / R}{y} = \gamma_L \frac{\vartheta I k_1^3}{s \chi_1 (l - l_1)} \tag{27}$$

$\kappa$ depends on the mode shape of the lever beam, the geometry, and the gauge factor.

The gauge factor $\gamma_L$ depends on the doping level and is between 10 and 100 [15].

## Circuit modeling

Figure 5 shows the equivalent small signal circuit of the ensemble variable resistance and its electronics. The expression of the transfer function $V_{out}/\Delta R$ is straightforward using the Millman's theorem at nodes 1 and 2,

$$V_1 \approx \frac{V_b \left( 2\delta R / R^2 \right) + V_2 ( j\omega C_{fb} )}{2/R + j\omega (C_{fb} + C_{in})} \tag{28}$$

and,

$$V_{out} \equiv V_2 = \frac{- g_m V_1 + j\omega C_{fb} V_1}{j\omega (C_{fb} + C_L) + gds + 1/R_L} \tag{29}$$

$\delta R$ ($\Delta Z$ in Fig.1) is the resistance variation around its nominal value R. $V_b$ is the read voltage applied on the gauges.

At this stage some assumptions on the relative values of $R_L$, $C_L$, $C_{fb}$, and $C_{in}$ have to be made. In particular, the input current must flow through the resistance $R$. In practice, typical resistance values of gauge obtained in this work are close to 3 kΩ. $C_{in}$ is close to 10 fF in a co-integration approach to 1 pF when a stand-alone circuit is used. It means that the assumption is true up to 150 MHz for the worst case. Moreover the output current must go through the output capacitance $C_L$. These conditions are summed up as,

$$1/\Omega_1 C_L << (R_L/(1+g_{ds}R_L)) \text{ and } g_m >> \Omega_1 C_{fb} \tag{30}$$

From (29) and (30), one can deduce:

$$V_{out} \equiv V_2 = \frac{-g_m V_1}{j\omega C_L} \tag{31}$$

Using (31) and (28), $G(\omega)$ is,

$$G(\omega) = \frac{V_{out}}{\delta R/R}(\omega) = \frac{-2g_m V_b}{j\omega(2C_L + C_{fb} R g_m)} \tag{32}$$

The minimal transconductance $g_m$ is defined from the oscillation conditions (3), and the expression (32),

$$g_m \approx \frac{\Omega_1^3 M_{eff} C_L}{V_{dc} J\kappa Q} = \frac{M_{eff} C_L \Omega_1^3 g^2 s\chi_1(l-l_1)}{QV_{dc} V_b \varepsilon_0 ea\gamma_L Ik_1^3 \vartheta} \tag{33}$$

As in the capacitive case, the conditions (30) can be simplified using (33) and expressing the conductance $g_{ds}$ as a linear function of $g_m$:

$$\frac{1}{R_L C_L \Omega_1} + \beta \frac{M_{eff}}{V_{dc} J\kappa Q}\Omega_1^2 << 1, \forall \Omega_1 \tag{34}$$

$$\text{and } \frac{C_{fb}}{\dfrac{2V_{dc} J\kappa Q}{M_{eff}\Omega_1^3 R}} << 1 \tag{35}$$

**Noise modelling**

The thermomechanical noise remains unchanged (see equation (18)). The noise from the gauge is the Johnson noise. Its expression in (m²/√Hz) is deduced from the equations (27) and (28),

$$S^{y_n}_{Johnson}(\omega) = \frac{4k_B TR}{\kappa^2 V_{dc}^2}\left(1 + \frac{C_{fb} R g_m}{2C_L}\right)^2 \tag{33}$$

where R is the sum of the gauge resistance and the lead resistance. The electronics noise in (m²/√Hz) is for a MOSFET in strong inversion,

$$S^{y_n}_{Electronics}(\omega) = \frac{8k_B T}{3g_m \kappa^2 V_{dc}^2}\left(1 + \frac{C_{fb} R g_m}{2C_L}\right)^2 \tag{34}$$

where $g_m$ is the transconductance according to the equation (32).

The model does not take into account the 1/f-noise due to low charge carrier density in the semiconductor material. However, we considered high doped silicon gauges (at least $10^{19}$ cm$^{-3}$) built with a typical CMOS process, which reduces the 1/f-noise.

Discussion on the system performance is done in the last part of the article.

## Performance of piezoresistive and capacitive oscillators

We simulate the two kinds of oscillators according to the resonant frequency of the NEMS. The frequency is changed from 1 MHz to 50 MHz by varying the beam width *w*. All other geometrical parameters are kept constant (Table I). To study in a proper way the capacitive and piezoresistive oscillators, both devices are settled in their optimal working point regime i.e based on their highest detection gain for each. For both oscillators, a monolithic integration is considered so that the parasitic capacitances are minimized [17]. The simulations are performed considering 0.35 µm-CMOS technology [2]. $C_{in}$, and $C_{fb}$ are parasitic components. $C_{fb}$ will reasonably not exceed 5 fF (around 0.13 fF/µm for the 0.35 µm -CMOS technology) and $C_{in}$ will be close to 10 fF.

The actuation voltages $V_{dc}$ is the same for the two oscillators. We assume a deflection $y$ =10 nm (~gap/10), well below than the non linear displacement limit and the dynamic pull-in [18]. For the piezoresistive detection, the readout voltage $V_j$ should be chosen as high as possible. In the same time, $V_j$ must be low enough to avoid the gauge breaking due to a large current flowing through [19]. For the silicon gauge of ~3 kΩ with a gauge factor of 100 for $10^{19}$ $cm^{-3}$ doping level [15], the voltage $V_j$ is set to 600 mV to keep the gauge temperature growth around 100 K. For the capacitive detection, the readout voltage is directly $V_{dc}$ and set to the maximum acceptable value for CMOS compatibility ($V_{dc}$ = 3.3 V*)*. Table II gives the electrical parameters used in. the model.

Notice that all characteristics driving the actuation gains are the same for the two oscillators. A priori, the only design parameters are $R_L$ and $C_L$ that should be adjusted for each mechanical resonant frequency to optimize the oscillation conditions. As shown in Tab. III, $R_L$ and $C_L$ have the same impact on the gain and phase conditions for each technique. These parameters are set at reasonable values so that $1/R_L C_L \Omega_1 << 1$ up to 100 MHz. Clearly, the upper working frequency limit of the two oscillators could be optimized by fine tuning of these parameters. But this study is out of the framework of this paper.

For the capacitive transduction, the oscillation conditions (see table III) are shown in Fig. 6. The phase conditions are only observed if the oscillation frequency is below ~5 MHZ. Notice that the equation (19) follows a 1/f law at low frequency and a quadratic variation at high frequency. Gains $J$ and $\kappa$ are independent of the frequency. The requested transconductance $g_m$ increases with the frequency. Stiffer mechanical structures demand larger energy level to reach self-oscillation. This simple law is well expressed by the equation (17). Over the usable frequency range, $g_m$ is always smaller than $2.10^{-5}$ S. This value is rather easily reached with a co-integrated circuit (with a single aforementioned 0.35 µm-MOSFET). Similar results are presented in Fig. 7 for the semiconductor gauges. This transduction principle is more effective to get self-oscillation compared to the capacitive one. The useful transconductance $g_m$ for a semiconductor gauge varies over the 7- 47 MHz frequency range from some $10^{-4}$ S to $5.10^{-4}$ S. Notice that the equation (34) follows a 1/f law at low frequency and becomes constant at high frequency. Gains $J$ and $\kappa$ change according to a $f^2$ law.

Finally, the figure 8 gives the theoretical mass resolution for the two detection schemes. Table IV sums up all noise sources used in the model. Over 5 MHz-10 MHz frequency range, the resolutions are quite similar close to 10 zg for a 1 Hz-bandwidth. For highest frequency up to 50 MHz, the expected resolution may reach 1 zg with the piezoresistive detection. These results are rather comparable with experimental mass performance already published [20] [21]. The difference between the two detection schemes (in particular the inflection point for the capacitive detection) is mainly due to the electronic noise. In particular, this noise is inversely proportional to the detection gain $\kappa$, which is constant with the capacitive detection and changes according to a $f^2$ law with the piezoresistive detection.

## Conclusion

We have done a theoretical comparison between two in-plane oscillating architectures based on piezoresistive and capacitive detections respectively. The sensors are embedded in a closed loop system with an electronic readout circuitry to define the mass-sensor. The overall measurement chains are studied through two specific models. This work provides a comprehensive theoretical method that is valid for any capacitive or piezoresistive principles. It is a useful methodology to design compact oscillator circuit around NEMS application.

Considering that the capacitive and PZR oscillators are set in their optimal working PZR detection can work at higher frequency and offers therefore better mass resolution. This trend can be extended to any NEMS.

Moreover silicon gauge can reach similar mass resolution than metallic gauge commonly used elsewhere [21]. Although the Johnson noise is increased with the resistance value, the signal improvement is much higher than the noise worsening. This conclusion completes the analysis most widely held that considers the metallic gauge as the best candidate for ultimate mass resolution. The silicon piezoresistive detection is an alternative choice for the mass sensing application if the ambient temperature is not too low (freeze-out effect of carrier density).

To conclude, this work paves the way toward the very large scale integration of the piezoresistive NEMS with its co-integrated electronics.

## Acknowledgement

Authors acknowledge financial support from the French Carnot - NEMS contract. We also thank M.L. Roukes for fruitful discussions.

**TAB. I. Typical values of the mass sensor – (see Figure 2 for the geometrical parameters)**

| ***l*** (µm) | ***w*** (µm) | $l_1$ (µm) | $l_j$ (µm) | $w_j$ (nm) | ***a*** (µm) | ***g*** (µm) | $f_1$ (MHz) | ***e*** (nm) | **Composition** |
|---|---|---|---|---|---|---|---|---|---|
| 5 | 0.3 | 0.75 | 0.5 | 80 | 4 | 0.1 | 21 | *160* | High doping ($10^{19}$ $cm^{-3}$) |

**TAB. II. Electrical parameters**

| | $C_L$ (fF) | $C_{fb}$ (fF) | $C_{in}$ (fF) | $C_0$ (fF) | $R_L$ (MΩ) | $V_{dc}$ (V) | $V_j$ (V) | $R$ (kΩ) | $\gamma_L$ | $J$ | $\kappa$ |
|---|---|---|---|---|---|---|---|---|---|---|---|
| capacitive | 5 | 500 | 10 | 0.1 | 1 | 3.3 | | | | $\frac{\varepsilon_0 eaV_{dc}}{g^2}$ | $\frac{\varepsilon_0 el^2\eta_1}{g^2\chi(l-l_1)}$ |
| PZR | | | | | | | 0.6 | 3 | 100 | | $\gamma_L \frac{\vartheta Ik_1^3}{s\chi_1(l-l_1)}$ |

**TAB. III. Oscillation conditions**

| | Gain | Phase |
|---|---|---|
| capacitive | $g_m > \frac{C_L (C_{fb} + C_{in} + C_0) M_{eff} \Omega_1^3}{V_{dc} J\kappa Q}$ | $\frac{1}{R_L C_L \Omega_1} + \frac{\beta \Omega_1^2 M_{eff} (C_{fb} + C_{in} + C_0)}{V_{dc} J\kappa Q} << 1, \forall \Omega_1$ <br> $\frac{C_{fb}}{\frac{V_{read} J\kappa Q}{M_{eff} \Omega_1^2}} << 1$ |
| PZR | $g_m > \frac{\Omega_1^3 M_{eff} C_L}{V_{dc} J\kappa Q}$ | $\frac{1}{R_L C_L \Omega_1} + \beta \frac{M_{eff}}{V_{dc} J\kappa Q} \Omega_1^2 << 1, \forall \Omega_1$ <br> $\frac{C_{fb}}{\frac{2 V_{dc} J\kappa Q}{M_{eff} \Omega_1^3 R}} << 1$ |

**TAB. IV. Main noise sources**

| | $S^{y_n}_{Thermomechanical}$ | $S^{y_n}_{Johnson}$ | $S^{y_n}_{Electronics}$ |
|---|---|---|---|
| capacitive | $\frac{4k_BTQ}{M_{eff}\Omega_1^3}$ | 0 | $\frac{1}{V_{dc}^2\lvert\kappa\rvert^2}\left((C_{fb}+C_{in}+C_0)^2+\frac{C_{fb}^2g_m^2}{\Omega_1^2C_L^2}\right)\left(\frac{8k_BT}{3g_m}\right)$ |
| PZR | $\frac{4k_BTQ}{M_{eff}\Omega_1^3}$ | $\frac{4k_BTR}{\kappa^2V_G^2}\left(1+\frac{C_{fb}Rg_m}{2C_L}\right)^2$ | $\frac{8k_BT}{3g_m\kappa^2V_G^2}\left(1+\frac{C_{fb}Rg_m}{2C_L}\right)^2$ |

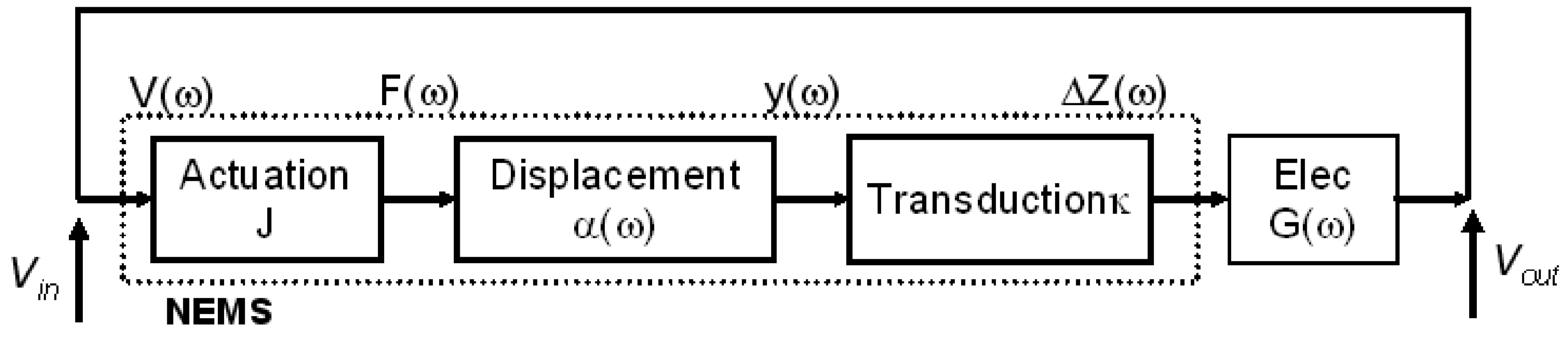

FIG.1. Closed loop representation of the self-exited system

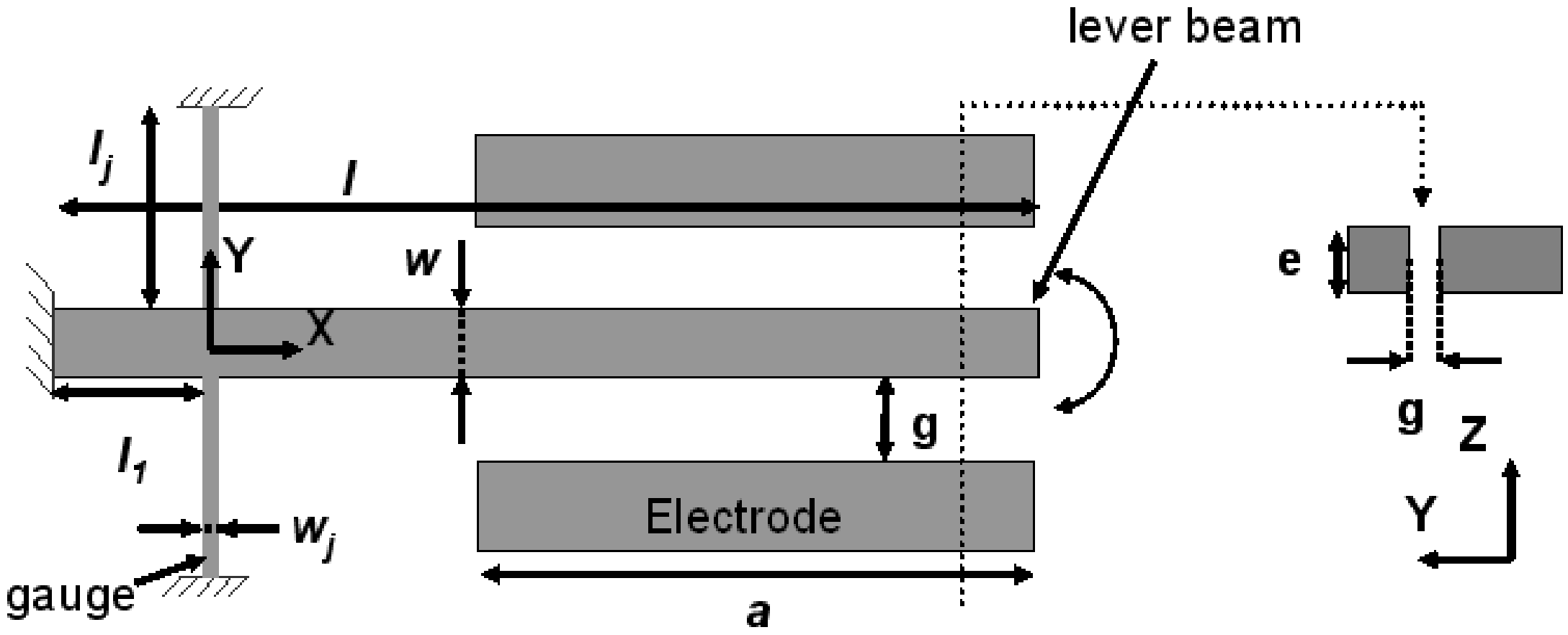

FIG. 2. Schematic of the cross beam used for the in-plane piezoresistive and capacitive detection

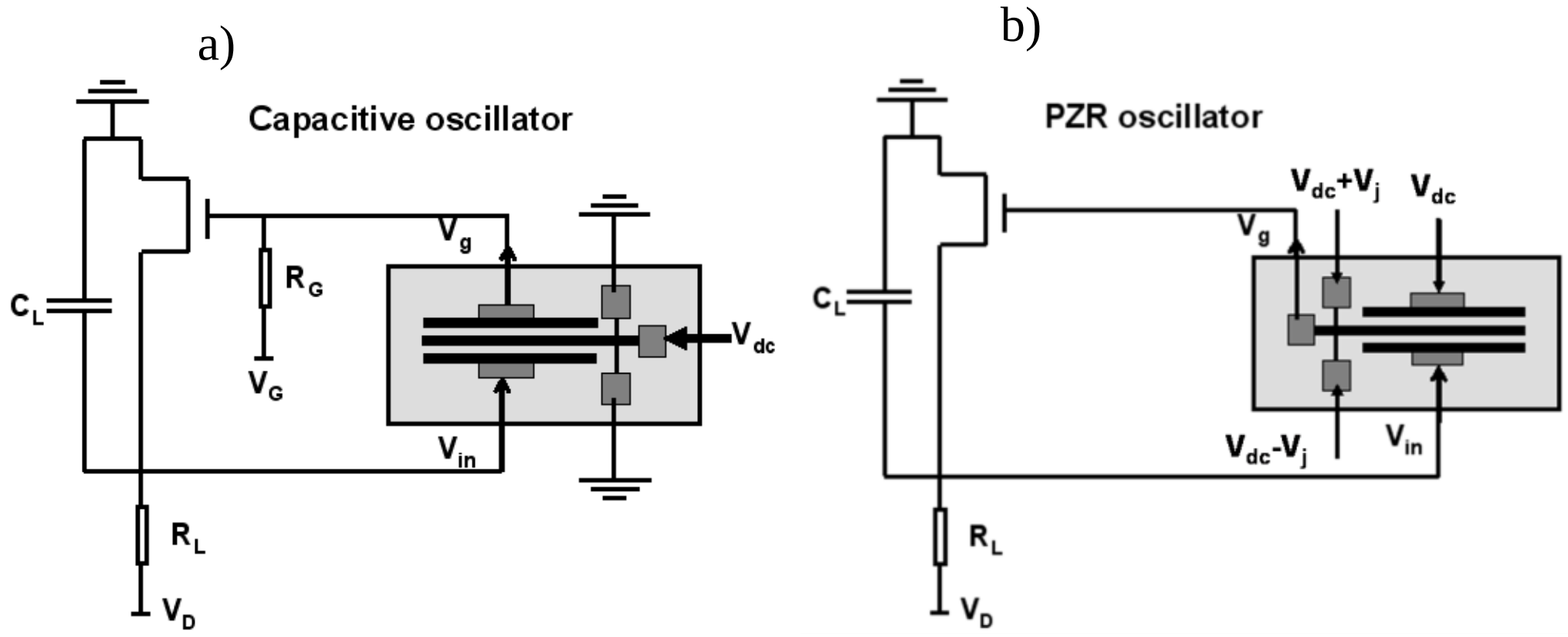

FIG. 3. Electrical schematic of the cross beam embedded into a self-oscillation loop – a) “Capacitive oscillator” scheme – b) “PZR oscillator” scheme – $V_G$ and $V_D$ are static polarization of the equivalent MOSFET – $R_G$ and $R_L$ are the polarization resistances - $C_L$

is the load capacitance – ($V_j$, $V_{dc}$) are readout voltages for the PZR and the capacitive detections respectively – $V_{dc}$ is the actuation voltage

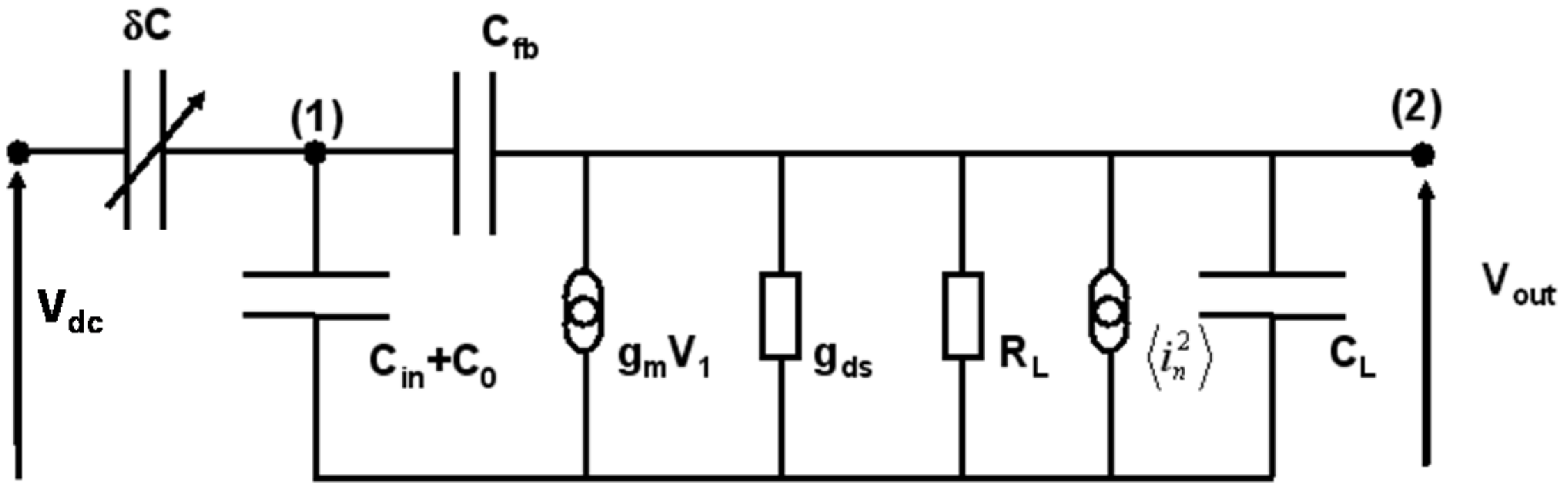

FIG. 4. Small signal model of the ensemble MOSFET+capacitive sensor, $C_{in}$ is the input capacitance, $C_0$ is the nominal NEMS sensor capacitance, $C_L$ is the output capacitance, $C_{fb}$ is the feedback capacitance, $g_m$ is the transconductance, $g_{ds}$ is the channel conductance, $R_L$ is the load resistance and $\langle i_n^2 \rangle$ is the noise current at the output of the MOSFET

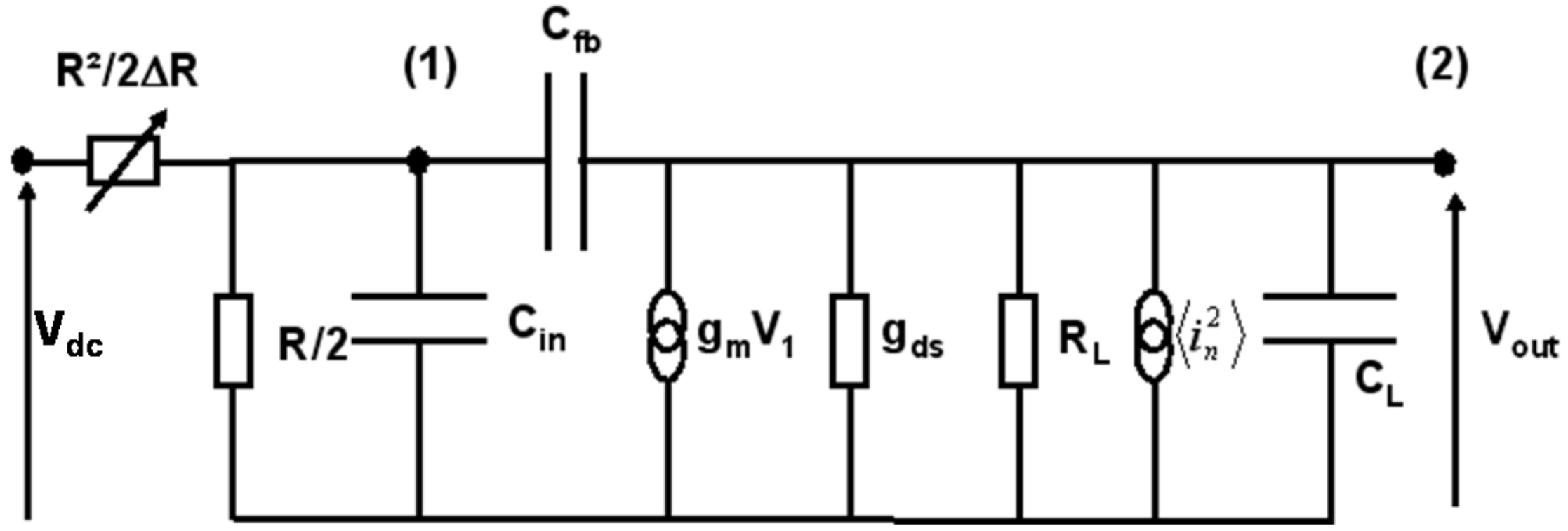

FIG. 5. Small signal model of the ensemble MOSFET+gauge, $C_{in}$ is the input capacitance, $C_L$ is the output capacitance, $C_{fb}$ is the feedback capacitance, $g_m$ is the transductance, $g_{ds}$ is the channel conductance, $R_L$ is the load resistance and $\langle i_n^2 \rangle$ is the noise current at the output of the MOSFET

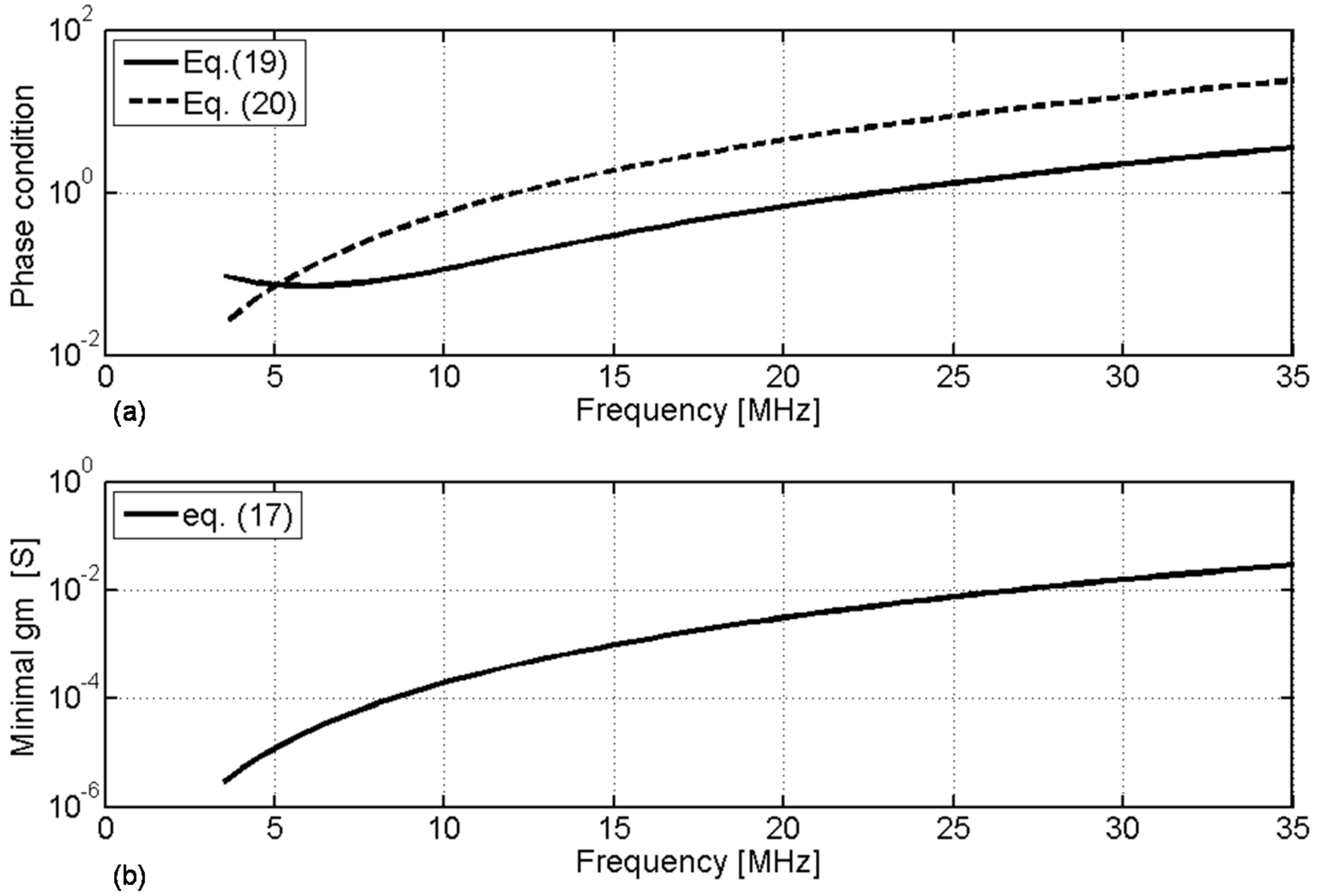

FIG. 6. (a) Phase conditions (19), (20) and (b) minimal transconductance (17) in the case of capacitive detection – Regions where (19) and (20) are higher than 0.1 are excluded to ensure the phase-conditions – $g_m$ should be lower than 1 mS to enable oscillations with a single MOSFET – Self-excited oscillation is performed up to 5 MHz.

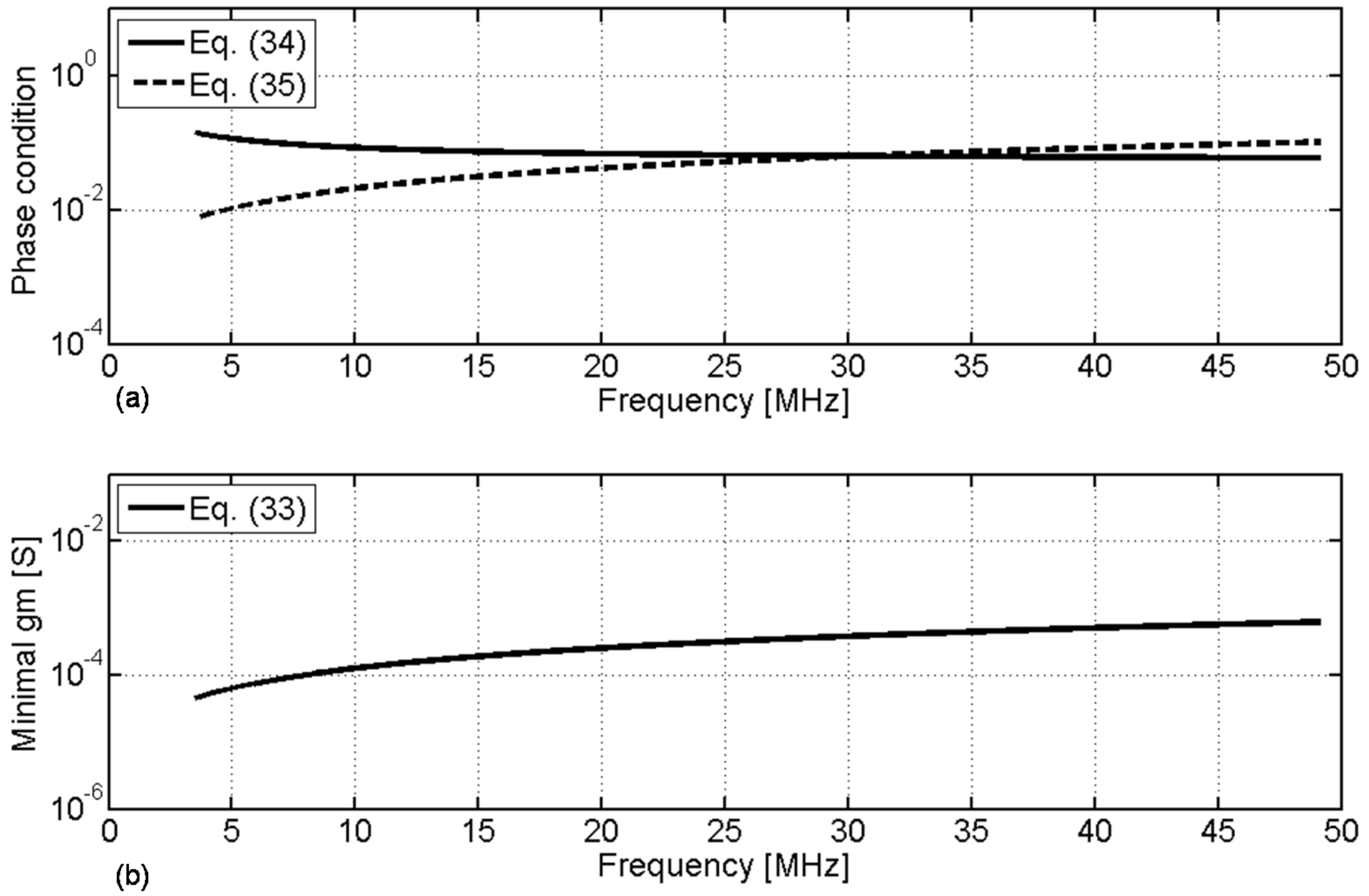

FIG. 7. (a) Phase conditions (34), (35) and (b) minimal transconductance (33) in the case of piezoresistive detection – Regions where (34) and (35) are higher than 0.1 are excluded to ensure the phase-conditions – $g_m$ should be lower than 1 mS to enable oscillations with a single MOSFET – The frequency range to get self-excited oscillation is 7 MHz to 47 MHz

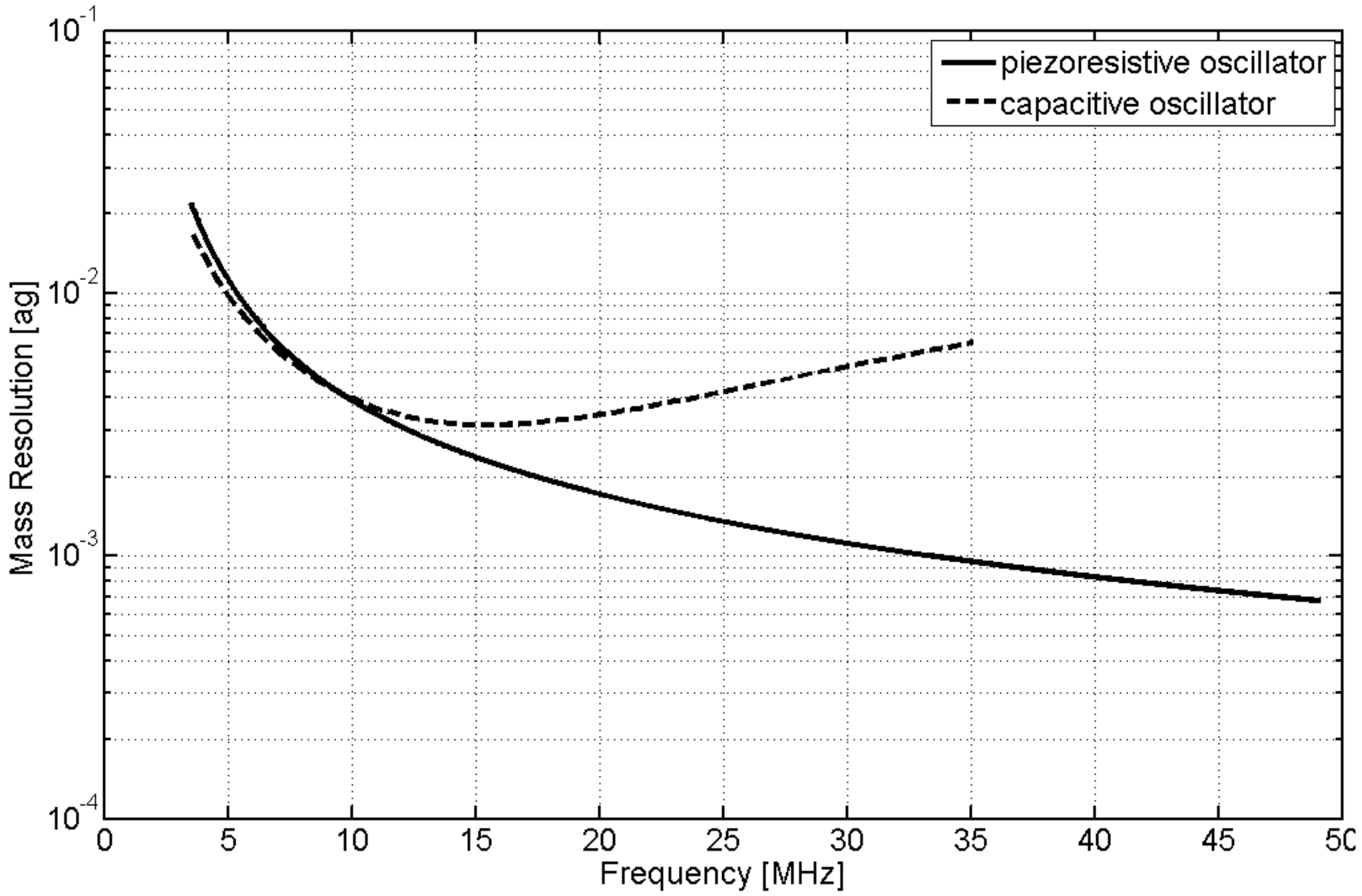

FIG. 8. Mass resolution of the oscillators.